\documentclass[%
twocolumn,
 reprint,
superscriptaddress,
groupedaddress,
 amsmath,amssymb,
 aps,
pra,
]{revtex4-2}

\usepackage{graphicx}
\usepackage{dcolumn}
\usepackage{bm}
\usepackage{wrapfig}
\usepackage{verbatim}
\usepackage{xcolor} 
\usepackage{soul}
\usepackage{adjustbox}

\usepackage{natbib}

\usepackage{hyperref}
\hypersetup{
        colorlinks=true,       
        linkcolor=blue,          
        citecolor=blue,        
        filecolor=magenta,      
        urlcolor=blue
        }

\begin{document}


\title{Rydberg Raman-Ramsey (R${}^3$) resonances in atomic vapor}

\author{Rob Behary} 
\affiliation{Dept. of Physics, William \& Mary, Williamsburg, VA 23187}
\author{Alex Gill}
\author{Aaron Buikema}
\affiliation{The Charles Stark Draper Laboratory, Inc., Cambridge, MA 02139}
\author{Eugeniy E. Mikhailov} 
\author{Irina Novikova} 
\affiliation{Dept. of Physics, William \& Mary, Williamsburg, VA 23187}


\date{\today}

\begin{abstract}

The sensitivity of electric field sensors based on two-photon electromagnetically induced transparency (EIT) involving highly excited Rydberg states in thermal atoms is often limited by the residual Doppler effect and optical power broadening. 
Here, we propose a method to reduce the EIT spectral linewidth using a Ramsey interrogation approach, allowing multiple interrogations of atomic coherence, using either temporally or spatially separated laser beams.  
Our theoretical calculations predict that the linewidth of such  Raman-Ramsey spectral features can be substantially reduced compare to a standard Doppler-broadened EIT, opening a possibility to improved sensitivity of Rydberg atomic vapor-based sensors. 
We also discuss some preliminary experimental efforts toward observing Raman-Ramsey resonances for both spatially and temporally separated interrogation approaches and associated technical challenges. 

\begin{description}
\item[PACS numbers]
\end{description}
\end{abstract}

\maketitle

\section{Introduction}
The application of atomic vapor cells for precision optical measurements and sensing has a long and successful history.
Electromagnetically induced transparency (EIT)~\cite{harris'97pt,lukin03rmp,FleischhauerRevModPhys05,Finkelstein_2023} enables a convenient link between optical transmission and minute variations in atomic energy levels caused by, e.g.,  external electromagnetic fields. 
Rydberg EIT in atomic vapors~\cite{MohapartaPRL2007,kubler_coherent_2010} has already become a well-established method for in-situ measurements of DC and rf electric fields~\cite{Raithel_PRA2019,Simons:18,gordonAIP2019,JauPRAppl2020,FancherIEEE2021}, development of broadband rf receivers and analyzers~\cite{CoxPRL2018,prajapati2022arxiv,MeyerPRappl2021}, THz imaging~\cite{hollowayAPL2014,PhysRevX.10.011027,Downes_2023}, SI-traceable electric field standards~\cite{HollowayJAP2017}, etc. 
However, power broadening and thermal atomic motion ultimately limit the width and amplitude of the observable EIT peak, deteriorating its achievable sensitivity. 

\begin{figure}[hbt!]
    \centering
    \includegraphics[width=0.9\columnwidth]{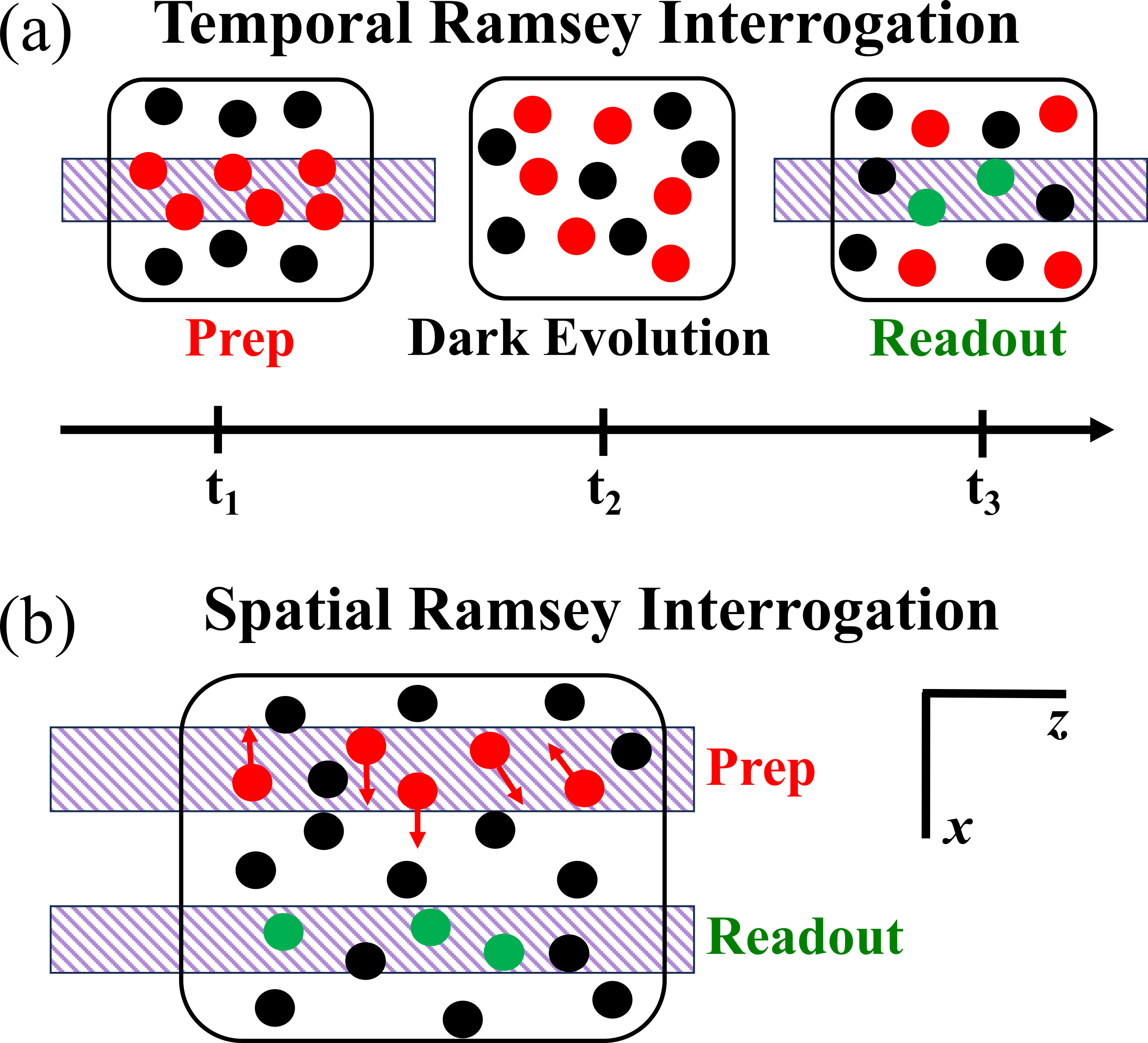}
    \caption{
        (a) \emph{Temporal R${}^3$ resonances:} At time $t = 0$ atoms are prepared in the coherent superposition with two EIT laser fields (red) for time $t_1$, and then allowed to evolve freely in the dark for time $t_2$. 
    Then, the changes in the atomic coherence are read out with a second light pulse (containing both EIT laser fields) of duration $t_3$. Only the initially prepared atoms remaining in the interaction volume (green) contribute to the R${}^{3}$ resonance.
    (b) \emph{Spatial R${}^3$ resonances:} Atoms are prepared in the coherent superposition with two EIT laser fields (red) in a prep interaction region, and move ballistically toward the readout region (also containing two EIT laser fields), crossing the dark region in between. 
    Only the atoms reaching the readout region (green)  contribute to the R${}^{3}$ resonance formation.
    }
    \label{fig:simplifiedExperimentPicture}
\end{figure}

In this paper, we theoretically investigate the possibility to overcome this limitation and obtain significantly narrower spectral resonances by using a Raman-Ramsey approach, in which the optical interrogation of atoms is interrupted by evolution in the dark. 
We consider two possible scenarios: temporal, which uses two bi-chromatic optical pulses shown in Fig.\ref{fig:simplifiedExperimentPicture}(a), and spatial, where moving atoms encounter two separate interaction regions shown in Fig.\ref{fig:simplifiedExperimentPicture}(b), created by CW optical fields. 
We show that in either case, it may be possible to produce a narrow Rydberg Raman-Ramsey (R${}^3$) resonance in addition to the standard Rydberg EIT peak. 
Since only slow atoms can constructively contribute to formation of R${}^3$ resonances,  the effect of the residual Doppler broadening can be significantly reduced. 
Our model predicts that R${}^3$ spectroscopy may provide superior sensitivity to the Rydberg level energy shifts, despite its reduced amplitude. 

In the following sections we present a theoretical model describing the R${}^3$ signal for two possible experimental configurations (time- and space-separated Ramsey interactions), and analyze the effect of atomic motion. 
Then, we describe experimental attempts to observe this effect and discuss technical issues that may hinder the observation of such resonances. 

\section{Theoretical Model}
\begin{figure}
    \centering
    \includegraphics[width=1\columnwidth]{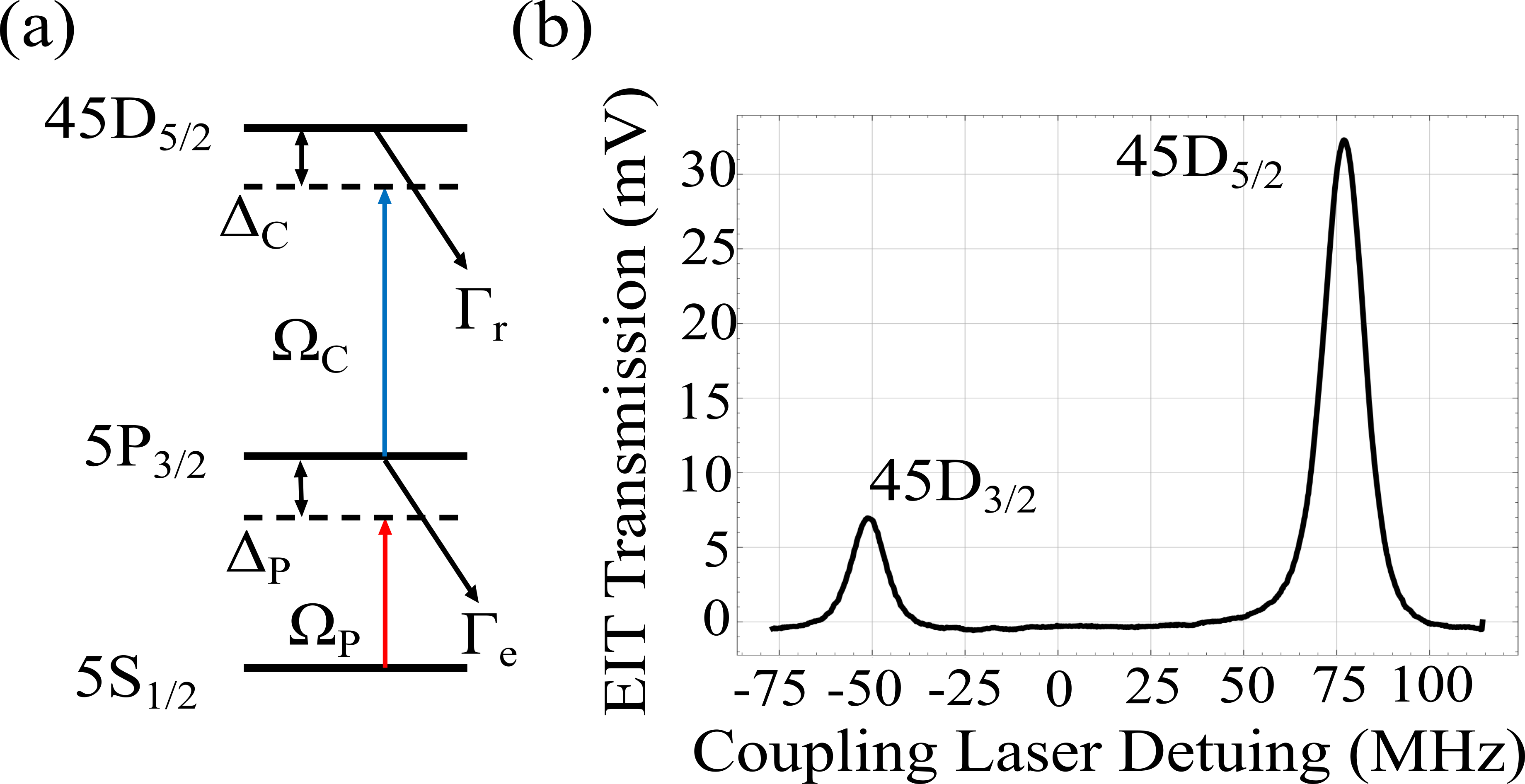}
    \caption{(a) Three-level ladder system with ground state $|g\rangle$, intermediate excited state $|e\rangle$ and Rydberg state $|r\rangle$ in ${}^{85}$Rb. 
    Probe laser with Rabi frequency $\Omega_P$ couples $|g\rangle$ and $|e\rangle$, while a coupling laser with Rabi frequency $\Omega_C$ couples $|e\rangle$ and $|r\rangle$. 
    $\Delta_P$ and $\Delta_C$ are the detunings of the probe and coupling lasers respectively. 
    (b) Experimental EIT spectra scanning the coupling laser detuning.}
    \label{fig:threeLevel}
\end{figure}

In this section we briefly summarize the three-level interaction model used to describe dynamic changes in the Rydberg EIT system shown in Fig. \ref{fig:threeLevel}(a) for stationary atoms. 
In this model, two optical laser fields (a \emph{probe} and a \emph{coupling}) interact with corresponding atomic optical transitions, as shown, and the transmission of the probe beam is measured.
The probe field with Rabi frequency $\Omega_P$ couples the ground state $|g\rangle$ with the first intermediate excited electron state $|e\rangle$, the population of which decays with a rate $\Gamma_e$ (in our system this is the $5S_{1/2}F=3\rightarrow 5P_{3/2}F'$ optical transition in ${}^{85}$Rb). 
Simultaneously, the coupling laser (Rabi frequency $\Omega_C$ couples state $|e\rangle$ to a Rydberg state $|r\rangle$ with significantly smaller decay rate $\Gamma_r$ (in our case $|r\rangle$ is the ${}^{85}$Rb $45D_{5/2}$ state 1/$\Gamma_r \approx 50 \mu s$). 
Using standard dipole and rotating wave approximations, the interaction Hamiltonian of the system can be written as~\cite{lukema,thesis2011,Scully}:

\begin{equation}
\label{eq:Hamiltonian}
    \hat{H} = \frac{\hbar}{2} \begin{pmatrix}
    0 & \Omega_P^* & 0\\
    \Omega_P & -2\Delta_P & \Omega_C^*\\
    0 & \Omega_C & -2(\Delta_P + \Delta_C)
    \end{pmatrix},
\end{equation}
where $\Delta_{P,C}$ are the one-photon detuning values of each laser from the corresponding optical transition. 
For now, we assume motionless atoms and discuss the effect of their thermal motion in the following sections.

To simulate the evolution of the atomic parameters,  we use the Lindblad master equation:
\begin{equation}
    \label{eq:Lindbland}
    \frac{\partial \rho}{\partial t} = -\frac{i}{\hbar}[\hat{H},\rho] + \sum_{i}\hat{\mathcal{L}}(\rho,\sigma_i),
\end{equation}
where, $\hat{\mathcal{L}}$ is the Lindblad superoperator given by $\mathcal{L}(\rho,\hat{\sigma}_i) = \sum_{i}\hat{\sigma}_{i}\rho\hat{\sigma}_i^\dagger - (\hat{\sigma}_i^\dagger\hat{\sigma}_{i}\rho + \rho\hat{\sigma}_i^\dagger\hat{\sigma}_{i})/2$, and $\hat{\sigma}_{i}$ are the collapse operators that account for the decoherence mechanisms.  
For our interaction system, we include the Lindblad superoperator $\hat{\mathcal{L}}_D$ for atomic decays which can be written as
\begin{equation}
    \label{eq:DephasingSuper}
    \hat{\mathcal{L}}_D = \begin{pmatrix}
    \Gamma_e \rho_{ee} & -\frac{\Gamma_e}{2}\rho_{ge} & -\frac{\Gamma_r}{2}\rho_{gr} \\
    -\frac{\Gamma_e}{2}\rho_{eg} & -\Gamma_e \rho_{ee} + \Gamma_r \rho_{rr} &-\frac{(\Gamma_e+\Gamma_r)}{2}\rho_{er} \\
    -\frac{\Gamma_r}{2}\rho_{rg} & -\frac{(\Gamma_e+\Gamma_r)}{2}\rho_{re} & -\Gamma_r \rho_{rr}
    \end{pmatrix},
\end{equation}
and the Lindblad superoperator $\hat{\mathcal{L}}_{lw}$ accounting for the finite width of the probe and coupling laser fields $\gamma_P$ and $\gamma_C$, correspondingly: 
\begin{equation}
    \label{eq:LinewidthSuper}
    \hat{\mathcal{L}}_{lw} = \begin{pmatrix}
    0 & \ -\gamma_P \rho_{ge} & -(\gamma_P + \gamma_C)\rho_{gr} \\
    -\gamma_P \rho_{eg} & 0 & -\gamma_C \rho_{er} \\
    -(\gamma_P + \gamma_C) \rho_{rg} & -\gamma_C \rho_{re} & 0
    \end{pmatrix}.
\end{equation}
Solving Eq.(\ref{eq:Lindbland}) for the matrix element $\rho_{ge}$, we find expressions for the complex optical susceptibility for the probe field, the absorption coefficient $\alpha$, and the refractive index $n$:
\begin{subequations}
    \label{eq:susceptibility}
    \begin{align}
    &\chi (\Delta_P, \Delta_C) = -\frac{2\mathcal{N}| \textbf{d}_{ge}  |^2}{\hbar \epsilon_0 \Omega_P} \rho_{eg}\\
    & \alpha = k_P\operatorname{Im}\left[\chi\right] \\
    & n = 1 + \operatorname{Re}\left[\chi\right]/2,
\end{align}
\end{subequations}
where $\mathcal{N}$ is the atomic  number density, $\textbf{d}_{ge}$ is the dipole moment of the $|g\rangle-|e\rangle$ atomic transition, and  $k_P = 2\pi / \lambda_P$ is the wave number of the probe optical field.

In the steady-state limit, one can easily reproduce the standard  EIT resonance in the probe field transmission, an experimental example of which is shown in Fig. \ref{fig:threeLevel}(b). 
In cold atoms, the width of such a peak is ultimately limited by the lifetime of the ground-Rydberg state coherence and can be very narrow (a few kHz). 
The width increases, however, with increasing laser power. 
For practical applications it is advantageous to use a somewhat power-broadened EIT resonance, as higher laser power provides more efficient coherence preparation and thus higher EIT resonance amplitude, as well as reduces the relative effect of the photon shot noise.

To reduce the effect of the power broadening without compromising the advantages of stronger probe optical field, we propose to use the  Raman-Ramsey interrogation scheme, consisting of two relatively strong laser pulses separated by a ``dark'' time, as shown in Fig. \ref{fig:simplifiedExperimentPicture}(a). 
In this case, atoms are first prepared in the desired coherent superposition of the ground and Rydberg atomic states using the first optical pulse, consisting of both EIT laser fields. Then the laser fields are quickly turned off, and 
the long-lived coherence between the ground and the Rydberg state $\rho_{gr}$ is allowed to evolve in the dark for time $1/\Gamma_r \ge t_2 \ge 1/\Gamma_e$. 
During this evolution its phase is affected by any perturbation in the Rydberg state energy that can be later read out by the second (detection) laser pulse whose power can be set with a sufficiently high signal-to-noise ratio. 
Most importantly, since most of the coherent evolution happens in the absence of interaction with optical fields, the effect of the power broadening is greatly reduced, even if strong fields are used in the preparation and detection stages. 
Such Raman-Ramsey interrogation schemes have been successfully implemented for improving the performance of the atomic clocks based on $\Lambda$-based EIT systems~\cite{zanonPRL2005,hafizJAP2017}, 

\begin{figure}
    \includegraphics[width=1\columnwidth]{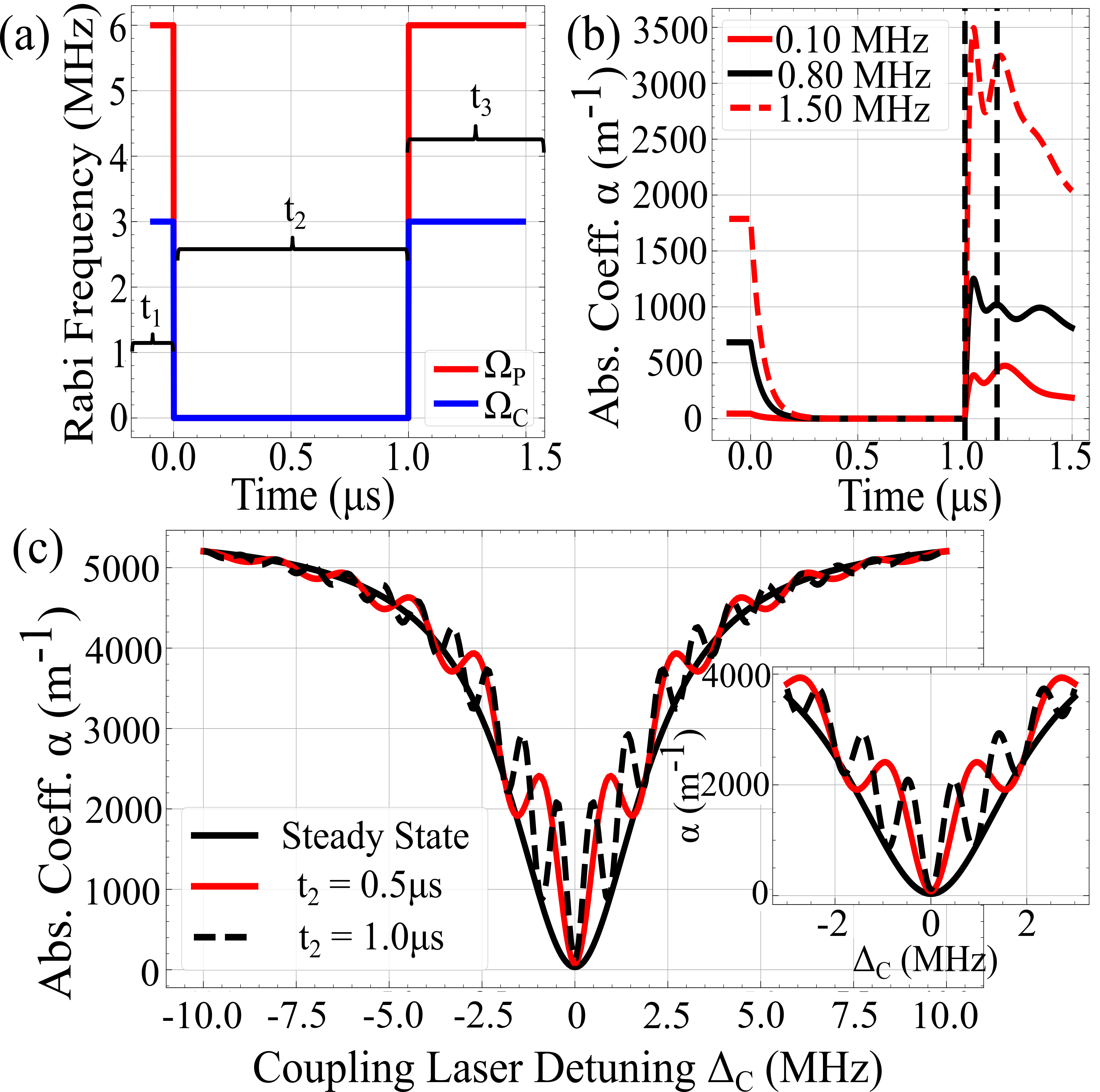}
    \caption{(a) Time sequence of temporal Ramsey interrogation for stationary atoms.  (b) Theoretically predicted R${}^{3}$ optical response for stationary atoms for different two-photon detunings $\delta_R$. 
    Black dashed lines indicate the $150$~ns integration region used to reconstruct Ramsey fringes. 
    (c) R${}^{3}$ resonances as a function of the coupling laser detuning for different evolution in the dark time $t_2$. 
    The model parameters are $\Delta_P$ = 0, $\Gamma_e$ = 6 MHz $\times$ 2$\pi$, $\Gamma_r$ = 3kHz $\times$ 2$\pi$, $\Omega_C$ = 0.5$\Gamma_e$, $\Omega_P$ = $\Gamma_e$, $\mathcal{N}$ = 1.7 $\times$ 10${}^{17}$ m${}^{-3}$, $\textbf{d}_{ge}$ = 1.46$\times$ 10${}^{-29}$ C$\cdot$m. 
    }
    \label{fig:RamseySingle}
\end{figure}


We calculate all atomic state evolution using the Python package QuTiP~\cite{qutip}, following the time sequence shown in Fig.\ref{fig:RamseySingle}(a). To simplify the calculations, we assume that the preparation time $t_1$ is long enough for the system to reach the steady state for the given initial laser parameters.
For the dark evolution time $t_2 \gg 1/\Gamma_e$,  the population of the intermediate excited state and the coherence between the ground state and intermediate excited state quickly vanish. 
On the other hand, the population of the Rydberg state $\rho_{rr}$ and the coherence between ground and Rydberg states $\rho_{gr}$ survive, as they decay much slower (this consideration can significantly simplify the numerical simulations by a simple analytical solution). 
For the remaining atomic coherence after the dark evolution time $t_2$, the analytical solutions are
\begin{subequations}
    \label{eq:density_after_dark}
    \begin{align}
    &\rho_{rr} = \rho_{rr,ss}e^{-t_2\Gamma_r}\\
    &\rho_{gg} = 1 - \rho_{rr} \\
    &\rho_{gr}  = \rho_{gr,ss}e^{-\frac{\Gamma_r}{2}t_2}e^{-i(\Delta_C + \Delta_P)t_2}.
\end{align}
\end{subequations}
In Eq (\ref{eq:density_after_dark}), $\rho_{rr,ss}$ and $ \rho_{gr,ss}$ refer to the steady state solution of the density matrix before the dark time.
In case of non-zero two-photon detuning $\delta_R=\Delta_C+\Delta_P \ne 0$, $\rho_{gr}$ acquires a phase $e^{i\delta_Rt_2}$. 
Finally, using the density matrix with only non-zero $\rho_{gg}$, $\rho_{rr}$, and $\rho_{gr}$ as the initial conditions, we numerically solve the Lindblad master equation in time for a fixed detection time $t_3$. 
The simulated time response of different coupling laser detunings is shown in Fig. \ref{fig:RamseySingle}(b).
Dotted vertical black lines in Fig. \ref{fig:RamseySingle}(b) show the region the atomic response is integrated over to produce R${}^{3}$ resonances. 
If this time is shorter than the time required to re-establish the steady-state EIT, the probe laser transmission is largely determined by the accumulated phase of $\rho_{gr}$, and displays a clear interference-like fringe pattern, as shown in Fig.~\ref{fig:RamseySingle}(c) with the period inversely proportional to the dark evolution time. 
Thus we can theoretically achieve the spectral resolution limited only by the Rydberg state decoherence time.

\section{Effect of atomic motion on the R${}^{3}$ resonances}

The motion of atoms in thermal atomic vapor greatly affects the characteristics of two-photon optical resonances. 
For temporal Raman-Ramsey excitation, we only consider atomic motion in the laser propagation direction, effectively assuming a large interaction volume (the transverse size of the laser beam in this case mainly limits the interaction time and effectively reduces the Rydberg state lifetime).  
The longitudinal motion along the laser propagation direction of atoms in the $+\hat{z}$-direction produces two generally undesirable effects on the R${}^{3}$ fringe formation, both related to the large frequency mismatch between two optical fields. 
The first one is the spatial phase variation between the probe and coupling fields, and the other is the differential Doppler shift for atoms with different longitudinal velocities. 

\begin{figure}
    \centering
    \includegraphics[width=1\columnwidth]{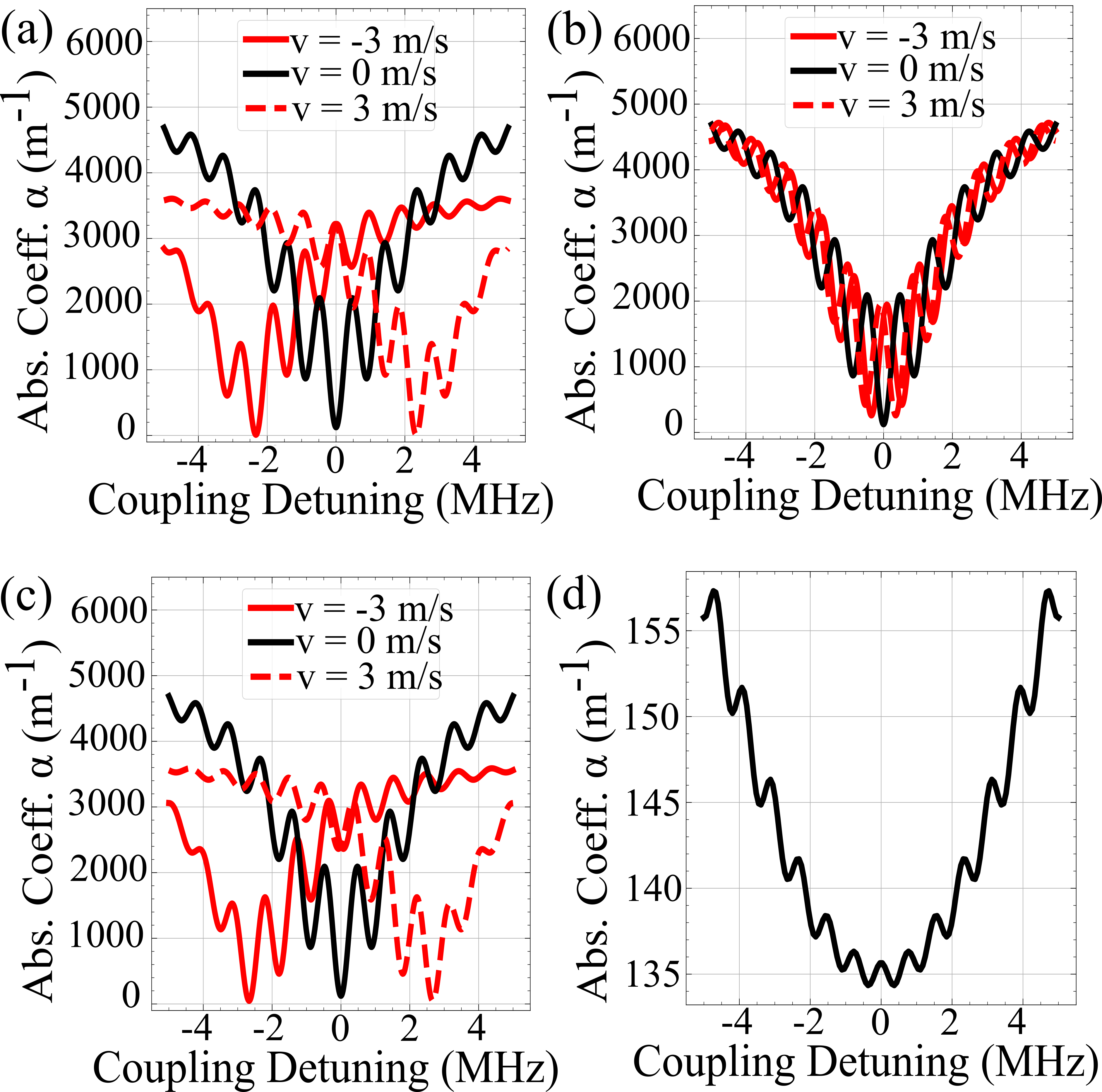}
    \caption{Examples of Ramsey fringe modifications for longitudinal motion of atoms ($\hat{z}$). 
    For (a) - (c) it is assumed that all atoms move with the given velocities $v_z$ to illustrate the effects of phase and Doppler mismatch. 
    (a) Effect of Doppler mismatch of detunings. 
    (b) Effect of spatial phase variations. 
    (c) Effect of both Doppler and phase variations. 
    (d) Fringes resulting from dark time $t_2$ = 1 $\mu$s and $t_3$ = 150 ns for a range of velocities integrated over a thermal 1D distribution corresponding to 300 K; note the different scale for $\alpha$.}
    \label{fig:longMotiononFringe}
\end{figure}

We consider the Doppler effect first, as it equally affects R${}^{3}$ and steady-state EIT resonances, and has been identified as one of the main factors limiting Rydberg EIT-based sensor sensitivity. 
If an atom moves with velocity $v_z$ along the laser beam, it ``sees'' the laser frequency $\omega$ shifted by $\omega v_z/c$. 
Even in the most beneficial geometry of the counter-propagating beams, there is a large velocity-dependent variation in the two-photon detuning $\delta_R(v_z)=(\Delta_C+\Delta_P-\omega_{rg}) + (k_P-k_C)v_z$. 
Practically, it means that even if the lasers are tuned precisely to the two-photon resonance, only a relatively small fraction of atoms with near-zero longitudinal velocities contribute to the EIT formation and that the observable EIT linewidth is broadened by this residual Doppler effect to a few MHz. 
In the case of Raman-Ramsey excitation, the Doppler effect causes the phase acquired during the dark evolution to depend on atomic velocity. 
This results in variations in the relative position of the fringes for each velocity, as shown in Fig.~\ref{fig:longMotiononFringe}(a). 
If this effect is considered in isolation, it can wash away the fringes entirely when the population is integrated over all velocity classes.

 We also need to take into account the spatial variation of the relative phase of the two optical fields along the beam path. 
 The initial phase of $\rho_{gr}$ is set by the relative phase of $\Omega_{C}$ and $\Omega_P$.
 Due to the mismatch in wavevectors, this phase is position dependent, so if an atom moves between the preparation and detection steps, the phase of the R${}^{3}$ fringe reflects this phase difference. 
 This effect almost exclusively affects R${}^3$ fringe formation, rather than steady-state EIT, since in the latter case the coherence phase adiabatically adjusts as atoms move along the laser beams. 
 However, in the Raman-Ramsey process, atoms detected at $z=0$  have been prepared at the location $-v_zt_2$, and thus their coherent state carries an additional phase $-(k_P-k_C)v_zt_2$ (for the counter-propagating optical fields). 
 Again, the contributions of different velocity groups destructively interfere, in principle limiting the fraction of atoms that can contribute to the observation of R${}^3$ resonance. 
 Fig.~\ref{fig:longMotiononFringe}(b) shows the fringe shifts due to the spatial phase mismatch for different $v_z$. 
 However, when considered together, these two phase shifts partially compensate each other. 
 Fig.\ref{fig:longMotiononFringe}(c) shows that the Doppler effect and relative displacement effects work together to bring the fringes back to constructive interference and preserve narrow fringes. 
 Thus, R${}^3$ fringes ``survive'' the integration of the 1D  Maxwell-Boltzmann velocity distribution of atoms in the $z$-direction, as shown in Fig.\ref{fig:longMotiononFringe}(d) for a single $t_2$ dark time, albeit with reduced amplitude compared to the cold atom case due to fewer participating atoms.  

This analysis gives a somewhat optimistic outlook on the possibility of observing narrower Rydberg EIT features using Raman-Ramsey excitation. 
However, from the practical point of view, such temporal interrogation may introduce some technical complications, such as the need for excellent pulse synchronization, fast pulse turn on/turn off, and a low detection duty cycle. 
An alternative approach for taking advantage of the same effect with steady-state lasers is implementing spatially separated preparation and detection regions. 
In this case, the dark time occurs naturally when atoms fly between the two interaction regions. 
Several geometries of such spatial multi-regional interaction have been tested previously for a $\Lambda$-type EIT systems~\cite{LezamaPhysRevA2010,RadojicicJOSAB15,jenkinsJPSAB19}.

\begin{figure}
    \centering
    \includegraphics[width=1\columnwidth]{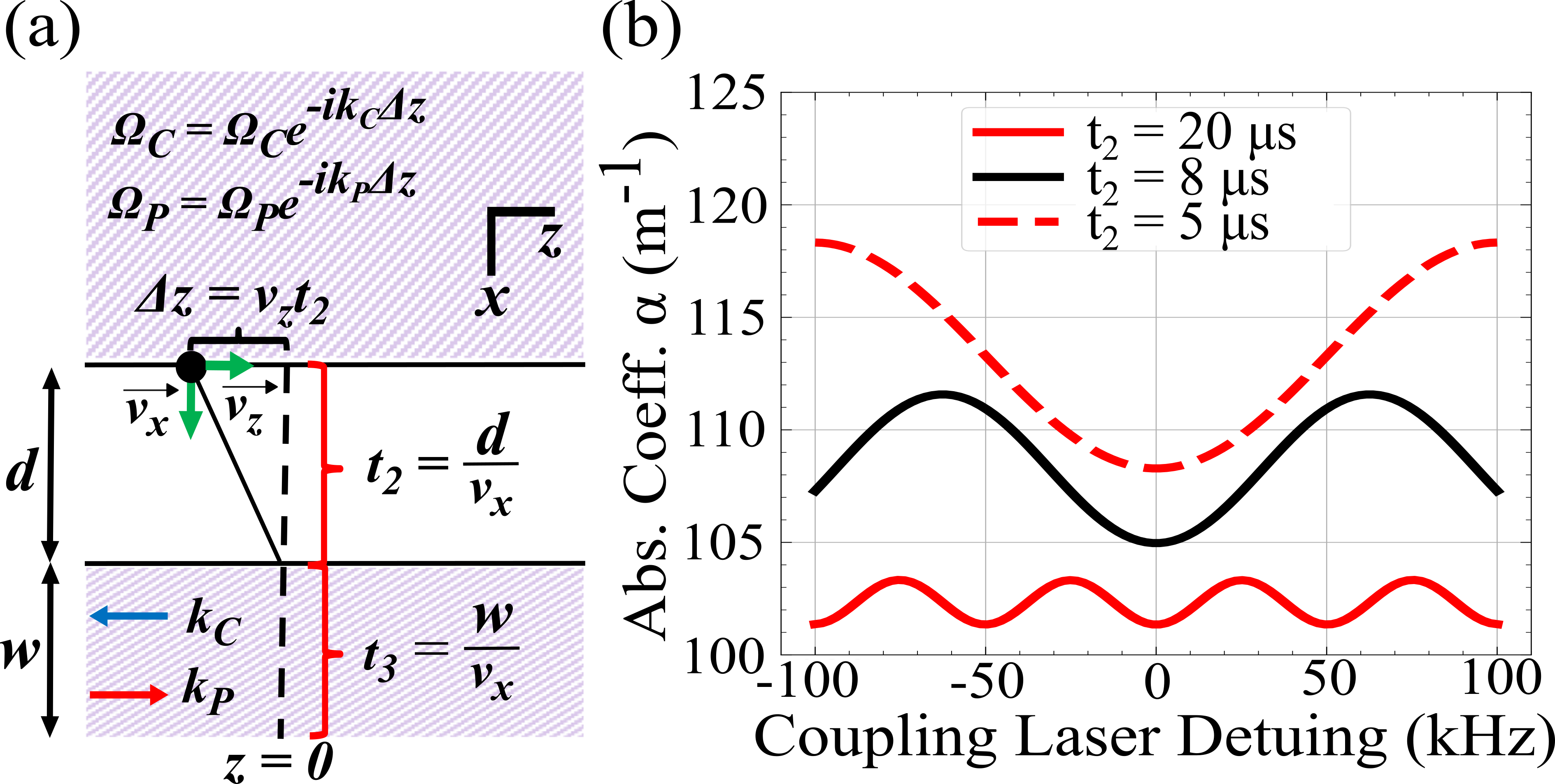}
    \caption{(a) Spatially separated geometry model. 
    Each interaction region (purple) contains counter-propagating probe and coupling beams. 
    The preparation region (top)  is assumed sufficiently wide, and separated by distance $d$ from the detection region that has a finite width $w$. 
    (b) Examples of Ramsey fringes for atoms with different transverse velocities $v_x$ (resulting in different effective dark times $t_2$) integrated over the longitudinal velocities $v_z$. }
    \label{fig:2DMotionExample}
\end{figure}

To properly describe the atomic response for the spatial version of R${}^3$ resonances,  we can assume two parallel interaction regions, separated by the gap, as shown in Fig.\ref{fig:simplifiedExperimentPicture}(b). To increase the number of interacting atoms, we assume that both regions are stretched in the perpendicular direction.
In this case, it is sufficient to consider the 2D motion of atoms between the two interaction regions as shown in Fig.~\ref{fig:2DMotionExample}(a). 
We assume that an atom first interacts with two laser fields in the preparation region long enough to reach the steady state. 
Atoms with nonzero transverse velocity $v_x$ eventually leave the preparation region and travel toward the detection region, effectively recreating the time sequence explored before and shown in  Fig.~\ref{fig:RamseySingle}(a). 
For each velocity group  $v_x$ the effective evolution in the dark time is $t_2=d/v_x$, where $d$ is the distance between the two interaction regions. 
As a result, atoms moving with different $x$-velocities produce Ramsey fringes with different periods, as shown in Fig.~\ref{fig:2DMotionExample}(b).
The fringes shown in Fig.~\ref{fig:2DMotionExample}(b) are also integrated over a range of thermal velocities in the $z$-direction in order to capture the effect of phase shifts due to relative displacement.
It will be shown that the central fringe survives when contributions of all velocity classes in both dimensions are integrated.

\begin{figure}
    \centering
    \includegraphics[width=0.75\columnwidth]{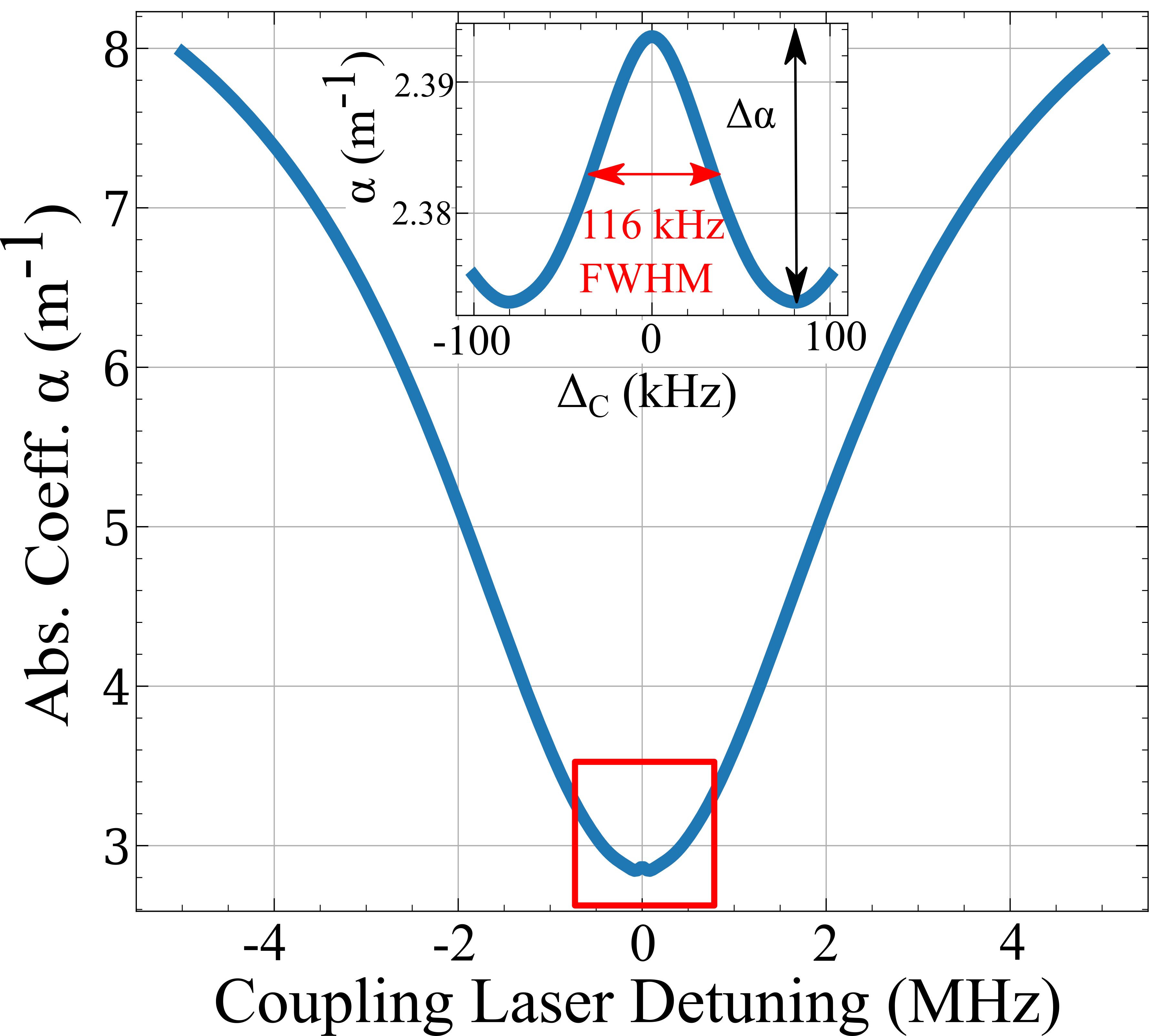}
    \caption{Predicted R${}^3$ absorption resonance as a function of coupling laser detuning for $d = $ 1 mm and $w = $ 50 $\mu$m. 
    The inset zooms into the narrow R${}^3$ feature.}
    \label{fig:R3_Resonance}
\end{figure}

Fig.~\ref{fig:R3_Resonance} shows a theoretically predicted R${}^3$ resonance, calculated using the same atomic parameters as Fig. \ref{fig:RamseySingle} with a beam separation of $d = $ 1~mm and a second interaction region $w = 50~\mu$m. 
Here we integrate the contributions to the optical susceptibility from each longitudinal velocity class for a given dark time $t_2$, and then integrate these results over all possible dark times $t_2=d/v_x$. 
Because of the mutually destructive contributions of atoms with different velocities, we carefully optimized number of velocity classes included in the simulations. 
For the longitudinal integration, the simulated density matrix element reaches a stable point at an integration range of -6$\mu\mathrm{m} \le\Delta z \le$ 6$\mu\mathrm{m}$ with 75 included velocity classes.
For the transverse integration range we use 0 $\le v_x \le$ 400 m/s with 50 velocity classes. 
These limits not only help manage simulation run time but also show the integrated signal is not a product of numerical instability.

The resulting optical absorption in Fig. \ref{fig:R3_Resonance} clearly shows two spectral features of different widths.  
The broader resonance is a ``standard'' Doppler-broadened EIT resonance due to atoms interacting with light only in the detection region. 
However, the second interaction for slower atoms produces an additional narrow feature on top of the broad EIT resonance - the R${}^3$ resonance. 
The details of this narrow spectral features are best shown in the inset of Fig. \ref{fig:R3_Resonance}. 
The full-width half maximum (FWHM) of the resulting resonance is quite narrow -  near 116 kHz, which is almost two orders of magnitude narrower than the standard EIT feature, even though its amplitude is much smaller as well.

\begin{figure}
    \centering
    \includegraphics[width=1\columnwidth]{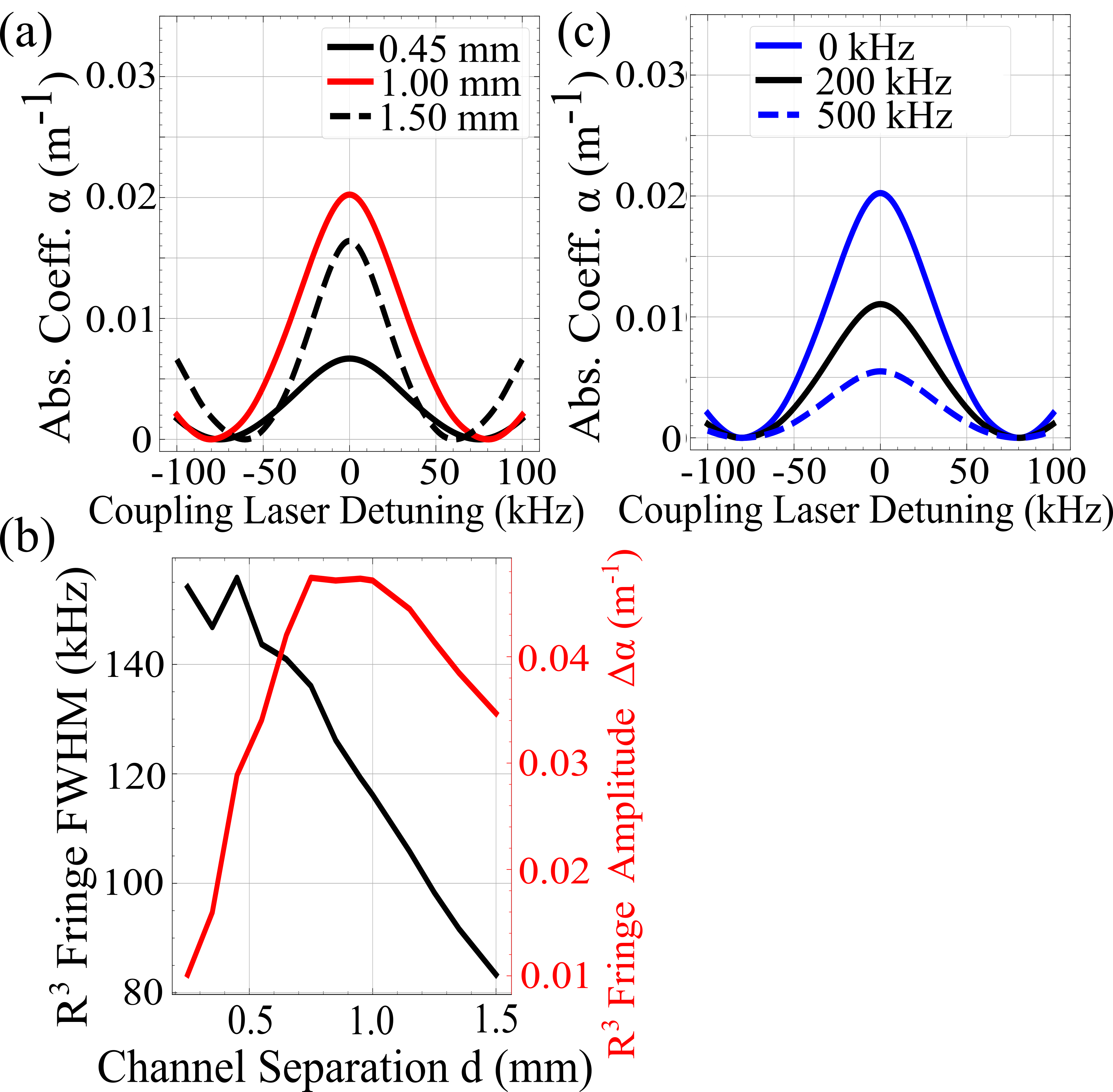}
    \caption{Simulated central R${}^{3}$ spatial fringes for varying measurement parameters. 
    (a) R${}^{3}$ fringe for various separation distances $d$.
    (b) Fit values for FWHM and height of the R${}^{3}$ fringe vs. $d$. 
    (c) Incorporation of 0 kHz, 200 kHz, and 500 kHz laser linewidths for the probe and coupling lasers. 
    Associated FWHM are 116 kHz, 121 kHz, and 125 kHz respectively.}
    \label{fig:R3_Perturbations}
\end{figure}

Fig.\ref{fig:R3_Perturbations} shows the impacts of both interaction separation distance $d$ and laser linewidths on the appearance of this central fringe.
Fig.\ref{fig:R3_Perturbations}(a) demonstrates the evolution of the fringe as the separation distance $d$ increases.
To quantitatively characterize the change in fringe parameters, we fit the calculated lineshapes using a quadratic background and Lorentzian peak to extract the FWHM and height of the R${}^3$ resonances.
This is plotted in Fig.\ref{fig:R3_Perturbations}(b) as a function of separation distance $d$. 
As expected, we observe a reduction of resonance width with larger separation; the height of the peak reaches its maximum value near $d = 1$~mm and then starts to decay, as fewer coherently prepared atoms survive the crossing between the two regions. 
Finally, Fig.\ref{fig:R3_Perturbations}(c) demonstrates the effect of the laser linewidths on the R${}^3$ resonance.
For this simulation the laser linewidths $\gamma_C$ and $\gamma_P$ are considered equal (for simplicity), and all the other parameters are the same as in Fig.\ref{fig:R3_Resonance}. 
The FWHM of the fringe remains relatively constant, but its amplitude drops as the laser linewidth increases, since this mechanism introduces additional dephasing and further reduces the fraction of atoms participating in R$^3$ resonance formation. 

\begin{figure}
    \centering
    \includegraphics[width=0.75\columnwidth]{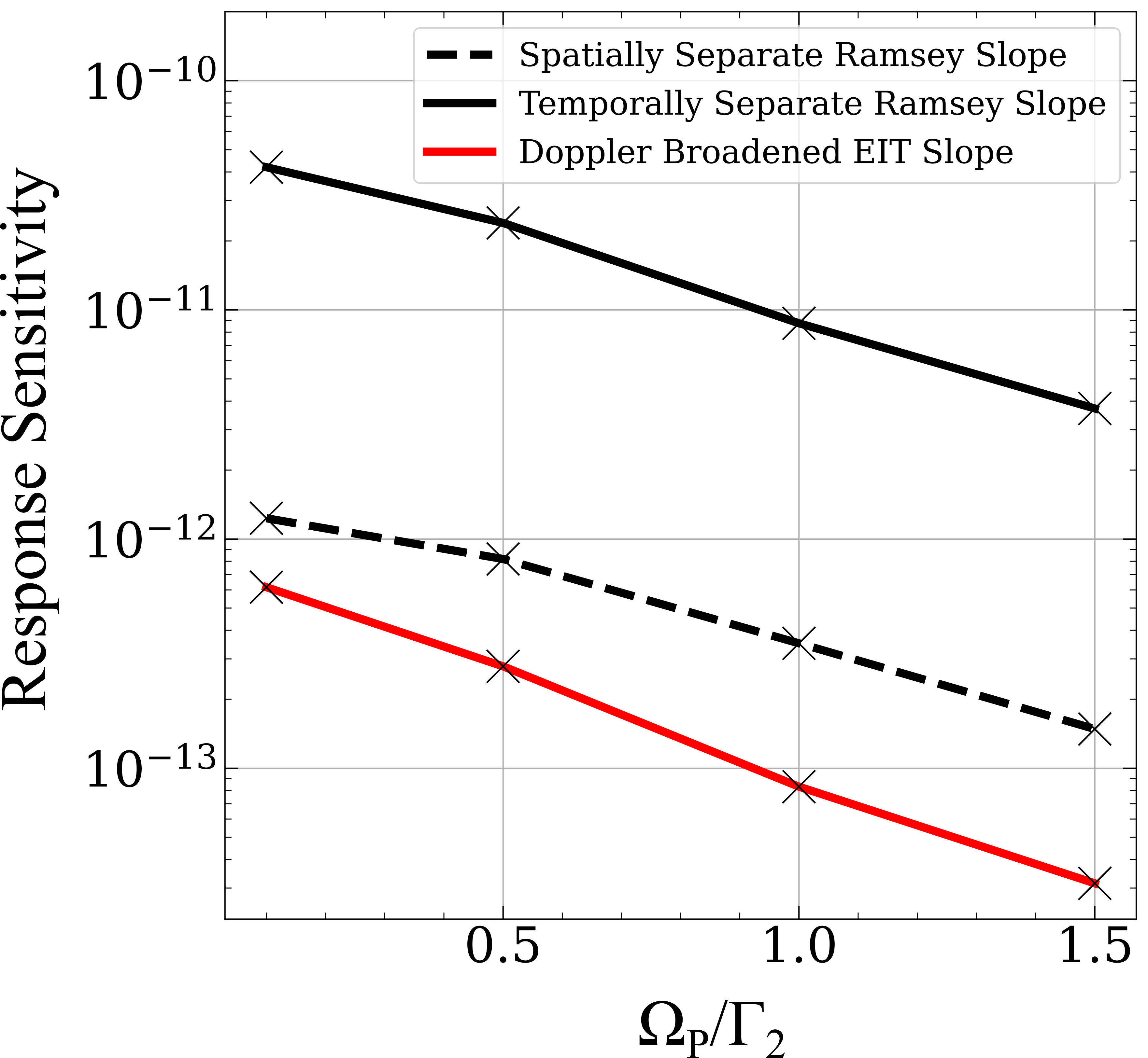}
    \caption{Comparison of the maximum slope near the central detuning $\Delta_C$ = 0. 
    The black x-marks indicate values of probe Rabi frequencies used, and the line behind them is a connecting line to show the trend. 
    Both the temporally and spatially separated cases display a similar trend of decreasing sensitivity for higher probe Rabi frequency.}
    \label{fig:EIT_comparison}
\end{figure}

To properly compare the potential advantage of a narrow R${}^3$ resonance over standard single-interaction steady-state EIT, we have to choose a figure of merit for spectral resolution (or sensitivity). 
Since typical optical sensors employ phase-sensitive detection, we can estimate the sensitivity by the first derivative of the spectral resonance near its center. 
We define sensitivity as the change in the absorption coefficient $\alpha$ over the change in coupling laser detuning $\Delta_C$. 
A comparison of the estimated sensitivity as a function of the probe Rabi frequency is shown in Fig. \ref{fig:EIT_comparison}. 
The predicted sensitivity values for the R${}^3$ resonance are consistently higher than for standard EIT over a wide range of probe intensities. 
Thus, at least in the idealized case considered here, the method of spatially and temporally separated Ramsey fringes can lead to a more sensitive measurement. 

\section{Preliminary steps toward Raman-Ramsey effect experimental realization}

The experimental realization of temporal or spatial R${}^3$ resonances imposes some reasonable yet challenging requirements on the laser system, particularly related to the relative phase stability between various optical fields. 
In the theoretical calculations above we assumed that all optical fields are phase-locked, so that the relative phase between probe and coupling optical fields evolve coherently. 
Any random phase change disrupts atomic coherence, resulting in reduced amplitude and increased spectral width of an EIT resonance. 
Consequently, in the case of the temporal Raman-Ramsey interrogation it is necessary to use lasers with coherence time longer than the dark time $t_2$ to ensure observation of the R${}^3$ fringes. 
Due to the large frequency difference between the two EIT lasers, it is impossible to phase-lock them electronically, even though it is possible to improve the relative frequency/phase stability by locking both lasers to the resonances of the same high-finesse cavity. 
The spatial realization adds an additional challenge of maintaining phase stability between the optical fields in the two interaction regions. 
Since it is natural to split the output of a single laser to produce either probe or pump beams in separate interaction regions, the phase stability requirement demands high-precision stabilization of the optical path length difference.  
While these technical challenges have known solutions, they require resources that were not available for the current efforts. 
As a result, we are not able to obtain the experimental verification of the reported theoretical results. 
Nevertheless, we observed some evidence of state preparation and transfer between two interrogations in both time- and space-separated cases.
 
\begin{figure}
    \centering
    \includegraphics[width=1\columnwidth]{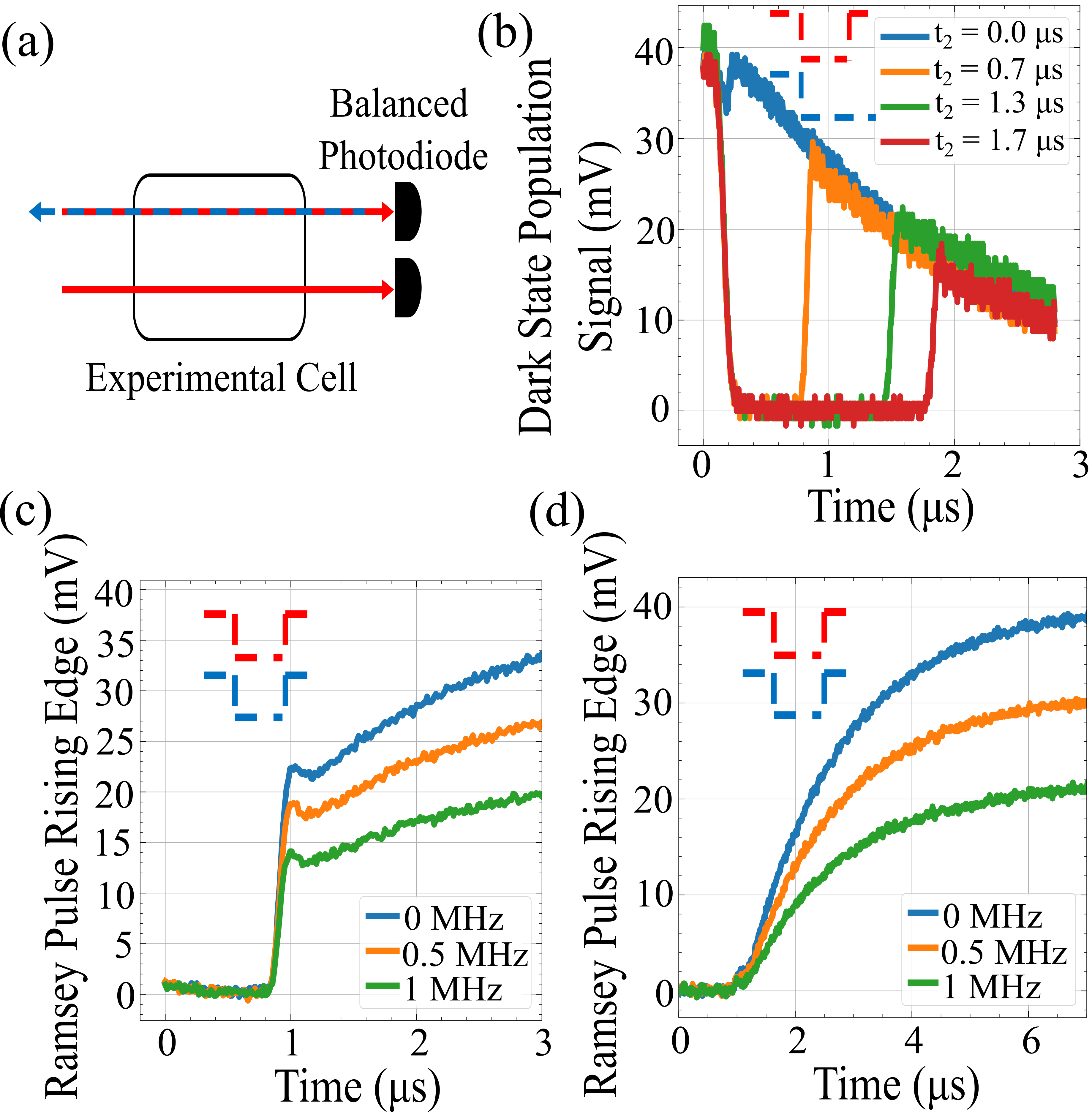}
    \caption{Experimental results for time-pulsed signals: (a) Balanced detection scheme used in the experiment. 
    (b) Probing dark state coherence in time.
    The coupling laser is not pulsed in these plots, and the probe laser is pulsed with the various dark times shown on the plot.
    While the lifetime of the state is ~50$\mu$s, the state decays in about 3 $\mu$s most likely due to atoms leaving the interaction region. 
    (c) and (d) are Ramsey style pulses as an attempt to see changes based on probe laser detuning in time.
    Both the probe and the coupling lasers are pulsed in these plots.
    (c) Short dark time $t_2$ = 1 $\mu$s where the dark state coherence is still present. (d) Long dark time $t_2$ = 8 $\mu$s showing the standard rise of EIT when the dark state coherence experimentally had decayed completely.}
    \label{fig:timeExp}
\end{figure}

For the experiments investigating the time-dependent transmission for the pulsed Raman-Ramsey effect, we used two counter-propagating laser beams interacting with a room temperature ${}^{85}$Rb vapor, and tuned to connect the ground state $5S_{1/2}F=3$ and Rydberg $45D_{5/2}$ state through the intermediate excited state $5P_{3/2}$. 
Both infrared probe ($\lambda_{p} = 780$~nm) and blue coupling ($\lambda_{c} = 480$~nm) lasers 
were intensity-modulated with two independent but synchronized acousto-optic modulators (AOMs), creating a sequence of 10~kHz square pulses with $\approx 200$~ns rise/fall fronts, separated by a variable $t_2$ dark time. 
A balanced detection scheme was implemented to detect exclusively two-photon-related transmission changes: we sent two identical infrared laser beams through the science cell, such that only one of them overlapped with the blue laser and then to a subtracting photodetector. 
Fig.~\ref{fig:timeExp}(a) shows a simplified version of the balanced detection scheme. 
Moreover, to remove the transient electronic feature in the first 200~ns of the rise and fall time, we also subtracted the differential signals in which only infrared beams were present in both channels. 
To realize that, the blue laser was present only every other time, so that we were able to record the sequence of two-laser and single-laser detector responses.
This is the acquisition method used for the plots in Fig.\ref{fig:timeExp}(b)-(d).

To investigate the dynamics of the Rydberg state population and coherence, we have conducted two types of measurements. 
One measurement probed the Rydberg state coherence lifetime, and the other attempted to measure the Ramsey interference.
In both measurements, the first (preparation) pulse $t_1$ time contained both infrared and blue optical fields to reach steady-state EIT conditions, followed by the ``dark'' time $t_2$, when both fields were off. 
In the first measurement shown in Fig.\ref{fig:timeExp}(b), only the probe laser was present for the second (detection) pulse $t_3$ time. 
For any $t_2 \gg 1/\Gamma_e$, the intermediate excited state population decays completely, such that changes in the transmission of the probe beam reflect variations in the ground state population. 
If some fraction of state populations is ``trapped'' in the Rydberg state, that reduces the ground state population, and the probe absorption is reduced. 
Fig.~\ref{fig:timeExp}(b) illustrates this behavior: for dark times shorter than the Rydberg state lifetime we see a gradual exponential transmission reduction toward its non-EIT level, likely due to a gradual decrease of atoms in the Rydberg state.
This result holds for multiple dark times.
The decay constant for this signal is $\approx 3~\mu$s, which is consistent with the average time atoms spend inside the interaction region. 
For these measurements, the two laser fields were tuned on the two-photon resonance, but variations of the two-photon detuning did not produce any noticeable changes in the transmission behavior, apart from overall rescaling. 
Since such measurements depend only on atomic population, this is expected.

When both laser fields are turned on during the detection pulse $t_3$ time, the rising edge of the pulse should ideally display an interference-like pattern, similar to those shown in Fig.~\ref{fig:RamseySingle}(b). 
For a fixed $t_2$ dark time, changes in the two-photon detuning $\delta_{R}$ should have a noticeable effect on the response when the light fields are turned back on in a Ramsey pulse before the atoms again reach the steady-state EIT condition. 
In the experiment, we adjusted the two-photon detuning by changing the probe AOM carrier frequency. 

Fig.~\ref{fig:timeExp}(c) shows the dynamics of the EIT signal for $t_2 = 1~\mu$s, where we expect a sizable percentage of coherently prepared atoms to be still present inside the interaction region. 
For comparison, we also observe the analogous signal for significantly longer dark time $t_2 = 8~\mu$s. 
In either case, we observe a rise in transmission likely due to atoms being repumped into the coherent EIT state (although the effect is more pronounced for longer dark time). 
The overall scaling of the detected transmission for different values of $\delta_{R}$ is consistent with the change in the steady-state EIT transmission levels. 
At the same time, we did not observe any signs of dynamic fringe detectable in the rising edge of the pulse. 

\begin{figure}
     \centering
     \includegraphics[width=0.75\columnwidth]{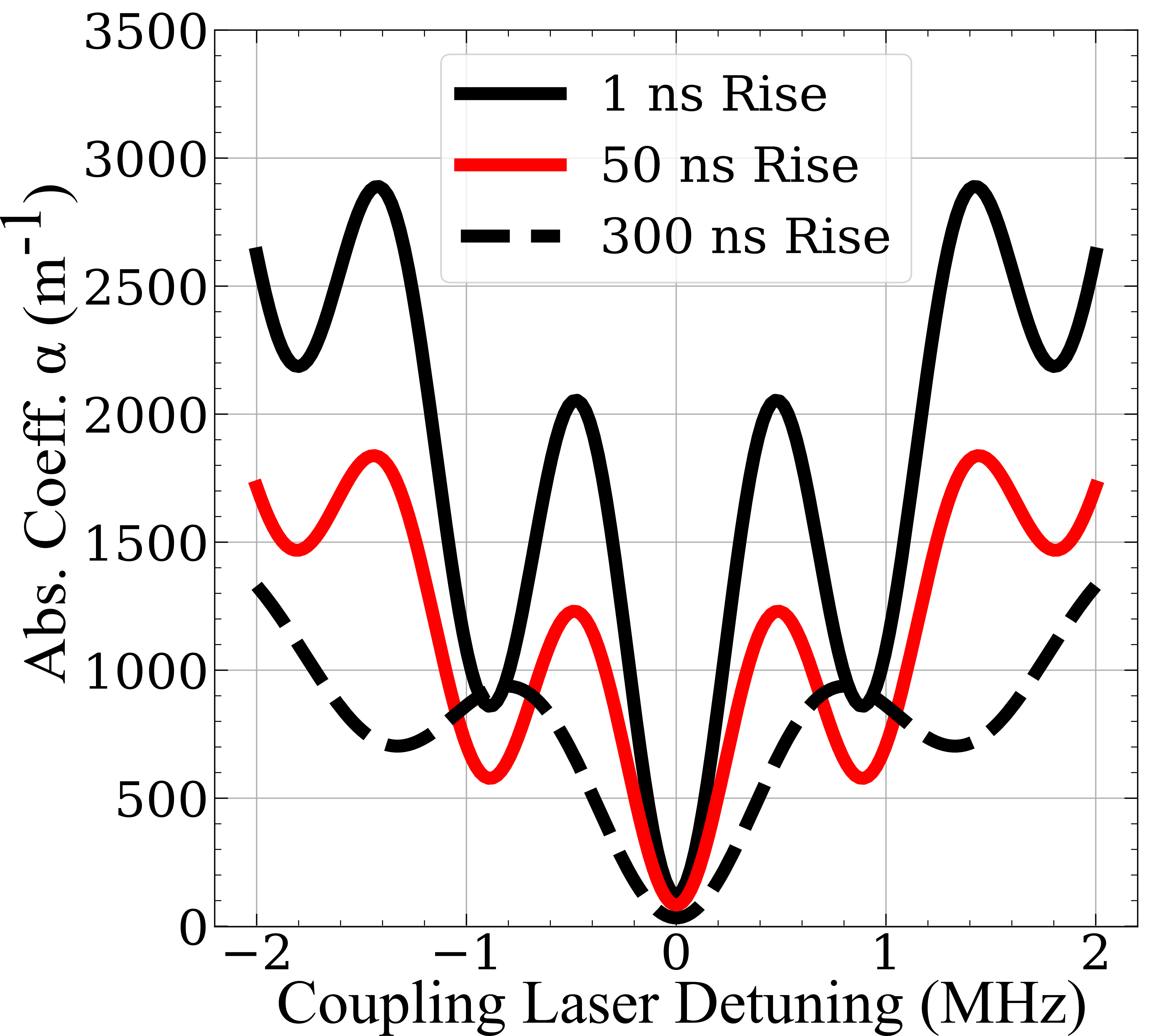}
     \caption{Simulated effect of slow rise time on Ramsey fringes for cold atoms. 
     All fringes are for $t_2$ = 1 $\mu$s and integration time $t_3$ = 150 ns from the start of the onset of the pulse. 
     For the shortest rise time of 1 ns (similar to the assumed instantaneous rise time of the model), the fringes show a higher contrast than for the slower turn on times.}
     \label{fig:slow_Rise_Effect}
\end{figure}

There are several technical explanations for why the observation of the resonances in our system may not have been successful. 
In addition to the relative phase stability issues mentioned above, our measurements were affected by the $150$~ns rise/fall times of the AOM switches, and potential instabilities of the relative turn on/off timing of the two AOMs. 
Fig.\ref{fig:slow_Rise_Effect} shows the effect that slow rise times have on the simulated Ramsey fringes for cold atoms. 
For the 200 ns rise time similar to what was seen in the experiment, the amplitude is half of the already small amplitude of the fringes for the near instantaneous 1 ns rise time. 
Therefore, it will be a necessity to have a near instant turn on time for the optical fields. 
We also suffered from some probe optical power and pointing fluctuations when adjusting the infrared AOM frequency. 
Finally, limited available blue laser power ($40$~mW) forced us to minimize the beam size (thus reducing the interaction time of atoms with the laser beam) and limited the achievable amplitude of the EIT resonance and, consequently, signal-to-noise ratio. 

\begin{figure}
    \centering
    \includegraphics[width=1\columnwidth]{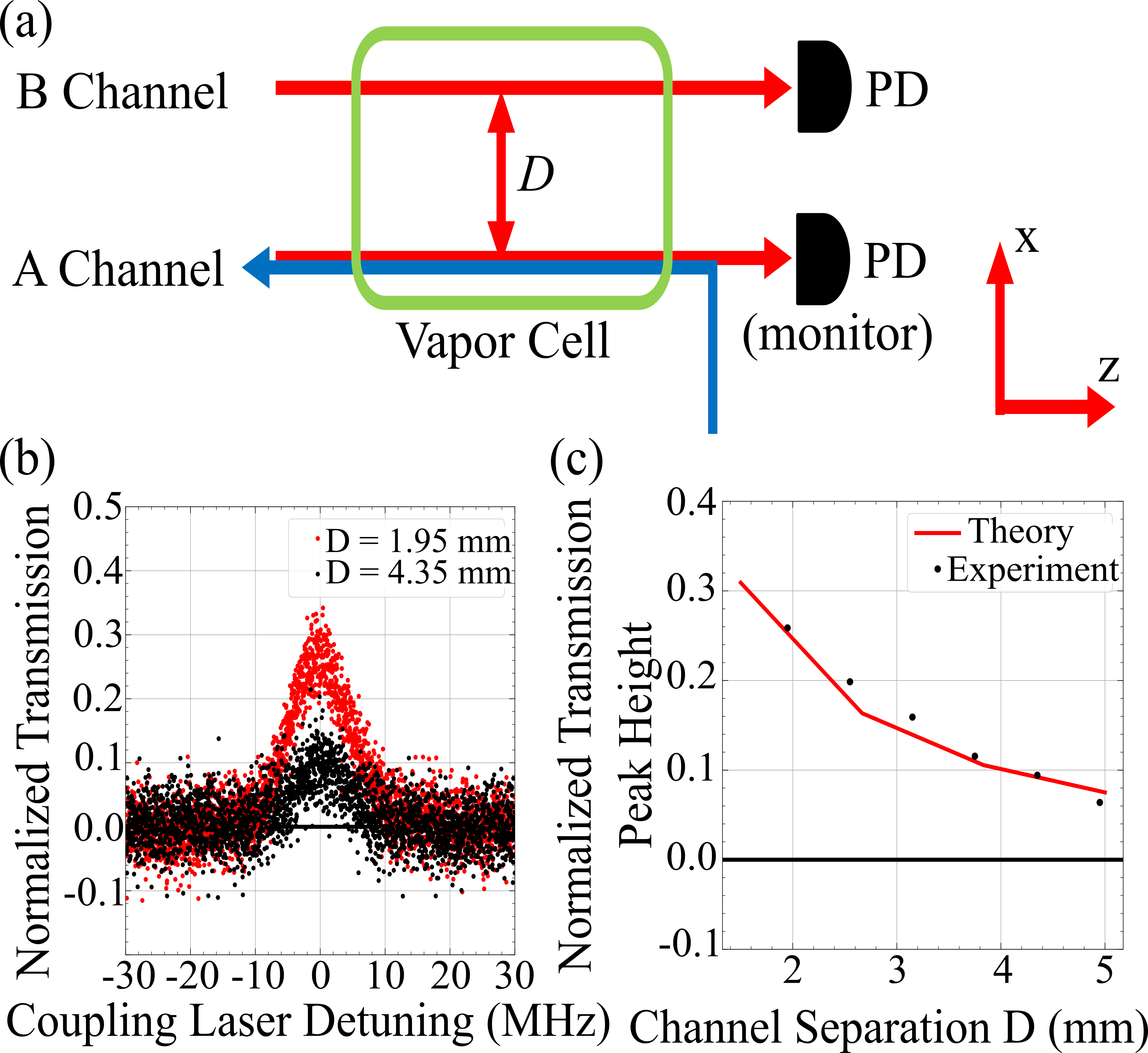}
    \caption{(a) Experimental setup for spatially separated beam geometry. (b) The transmission of infrared light in channel B for a given beam separation $D$. Reported transmission is normalized to the peak steady-state EIT transmission experimentally observed using only the infrared and blue beams of channel B (channel A blocked).(c) The experimentally measured EIT peak height as a function of beam separation, as compared with the results predicted by our theoretical model. There are no free parameters in our theoretical result. The validity of our model is supported by the close correspondence between theoretical and experimental transmission curves.}
    \label{fig:spatialExperiment}
\end{figure}

For the experiments investigating the spatially separated beam geometry for the detection of the R${}^3$ resonance, we again use a two channel detection scheme as shown in Fig.\ref{fig:spatialExperiment} (a). 
Channels A and B refer to the two spatially separated optical interrogation regions, each containing one or both of the 780 nm probe or 480 nm coupling light propagating along the $z$-direction. 
The two photodiodes are monitored independently to measure changes in the absorption of either probe beam.
For these experiments care is taken to avoid optical crosstalk between the channels.
The beam diameter of each beam in the $x$-dimension is approximately 0.1 mm.
The onset of optical interference between beams of the same color occurs for separation distances below $D$ $\approx$ 0.5 mm.
This is confirmed using a beam profiler placed at the position of the cell.
Sufficient channel separation must be maintained in order to eliminate the effects of crosstalk between channels. 
We ensure, for instance, that there is negligible channel A probe light striking the channel B detector and vice versa.
We also ensure that there is negligible overlap of channel A coupling light with the channel B probe, as evidenced by the lack of an observable EIT signal seen from just these two fields alone.
The separation distance $D$ can be manually adjusted to explore the effect of this parameter on photodiode signals.
For these experiments, our coupling laser is tuned to resonance with the $5P_{3/2}\rightarrow 56D_{5/2}$ transition.


Using this setup we successfully observed the effects of the flight of Rydberg atoms between the two interaction regions. 
This is analogous to the temporal probing of Rydberg population after a variable dark time shown in Fig~\ref{fig:timeExp}(b). 
In this experiment, channel A prepares the atoms in the steady state EIT condition, and Channel B detects the atomic signal using only the 780 nm probe laser. 
Deviations from the nominal probe absorption of the Doppler-broadened 5$S$ to 5$P$ transition observed by the channel B probe result from the presence of atoms prepared in a Rydberg-dressed dark state in channel A.
Adjusting the intra-channel separation D is analogous to adjusting the dark time in our experiments with temporally pulsed laser beams.
In Fig.\ref{fig:spatialExperiment}(b) we show that as the channels are moved further apart the detected EIT peak in channel B decreases in amplitude.
This result is supported by the numerical model and is qualitatively explained by the correlation of larger values of $D$ with higher average transverse speed $v_x$ of the detected Rydberg atom population.
Because the readout channel has a fixed width, a higher transverse speed results in a reduced readout interrogation time and a decrease in signal amplitude.

While this observation of the flight of room temperature Rydberg-dressed atoms between two channels is a necessary first step for spatially-separated Ramsey interrogation, the setup is not yet optimized to achieve sufficient signal-to-noise ratio to observe a Ramsey fringe experimentally. 
Additionally, the aforementioned laser frequency and optical path-length stabilization concerns must be addressed in order to stabilize the anticipated Ramsey phase.

\section{Summary}

In this paper we develop a model for Rydberg Raman-Ramsey (R${}^3$) EIT resonances by calculating the atomic response due to the repeated interactions with two optical fields in two-photon Raman resonances. We focused on the effect of the atomic motion on the resulting narrow spectral feature for both spatially or temporally separated interaction regions. 
%
%
We predict that even for thermal atoms it should be possible to observe R${}^3$, and that using this resonance can potentially improve sensitivity to electric field measurements. 
Experimental efforts to observe R${}^3$ resonances in both the temporal and spatial configurations of the experiment were not successful, likely due to insufficient relative phase stability and imperfect pulse turn on and off. At the same time, we observe characteristic change in the optical response due to the atomic preparation in the Rydberg state. 

Sponsored by: Defense Advanced Research Projects Agency\\
DARPA Program: BAA HR001120S0048 - ``Defense Sciences Office Office-wide''\\
Issued by DARPA/CMO under Contract NO. HR0011-22-C-0061

\bibliographystyle{apsrev4-1}
\bibliography{bibliography,bibliographyIN} 

\end{document}